\documentclass[aps,amsmath,amssymb,showpacs]{revtex4}

\begin{document}

\title{ Measuring Unrecorded Measurement.}

\author{M. Revzen and A. Mann}
\affiliation {Department of Physics, Technion - Israel Institute of Technology,
Haifa 32000, Israel}

\date{\today}

\begin{abstract}

Projective (Von Neumann) Measurement of an operator (i.e. a dynamical
variable) selected from a prescribed set of operators is termed unrecorded
measurement (URM) when both the selected operator and the measurement outcome
are unknown, i.e. "lost". Within classical physics a URM is completely
inconsequential: the state is unaffected by measurement. Within quantum
physics a measurement leaves a mark. The present study provides
protocols that allow retrieval of some of the data lost in a URM.\\

\end{abstract}


\maketitle

\section{Introduction}

Within quantum mechanics a  measurement (the present  study is confined to Von
Neumann, i.e. projective, measurements, \cite{barnett,nielsen}) has important
effect. This remains true for a measurement of an operator, selected from a
prescribed set of operators, with neither the measurement outcome nor the
operator chosen being available, i.e. are "lost". We refer to such measurement
as "unrecorded measurement" (URM). E.g., confining ourselves to a two
dimensional Hilbert space, with measurement of the spin along the x direction
for a particle whose spin is aligned along the y direction. The measurement
outcome has equal chance of being $+1$ or $-1$. Let the outcome be $+1$. This
is a URM measurement if neither the direction (here x) nor the outcome (here
+1) are available ( are "lost"). Nonetheless the state of system has changed:
the particle did enjoy a (Von Neumann) measurement. QM, \cite{peres}, assures
us that should we measure the particle's spin along the x direction the result
would be $+1$. We seek a protocol prescribing  "control measurement" that will
disclose (some) of the changes in the system that underwent a URM. The study
below considers setups allowing  retrieval of (some of) the unavailable data
of a URM. We note that these are purely quantum in nature as within the
classical theory URM
is completely inconsequential and leaves, in principle,  no trace.\\

Our strategy is to entangle our system (particle "1" - the system that will
undergo a URM) with an ancilla (particle "2") \cite{barnett}. The URM, though
pertaining to particle 1, affects both systems and we shall extract the
information we seek from both. This is achieved by a two-particles
control measurement of the combined system.\\
The (single particle) URM bases that we consider for a d-dimensional Hilbert
space particle are mutually unbiased bases, MUB. To assure self containment
and to fix the notation we now give a brief review of MUB
\cite{schwinger,w1,w2,w3,tal,vourdas,bg1,
klimov,ivanovich,r1,r2}.\\

 Two orthonormal vectorial bases ${\cal{B}}_1,\;{\cal{B}}_2$  are said to be
MUB if and only if $({\cal{B}}_1 \ne {\cal{B}}_2)$
\begin{equation}
\forall\;|u\rangle,\;|v\rangle \in {\cal{B}}_1,\;{\cal{B}}_2\;resp.\;|\langle
u|v\rangle|=\frac{1}{\sqrt d}.
\end{equation}
A set of orthonormal bases which are pairwise MUB is a MUB set. It was shown
in \cite{ivanovich} that there are at most d+1 MUB in a set belonging to a
d-dimensional space. For d=prime ($d\ne 2$), d members of an MUB set are given
in terms of the (d+1)-th basis $\{|n\rangle\}, n=0,1,...d-1$ by (b designates
a basis, m specifies the vector in the basis)
\begin{equation}\label{mub}
|m;b\rangle=\frac{1}{\sqrt
d}\sum_{n=0}^{d-1}|n\rangle\omega^{\frac{b}{2}[n(n-1)]-mn};\;b=0,1,...d-1;\;\;\omega=e^{i\frac{2\pi}{d}}.
\end{equation}
The (d+1)-th basis, termed the computational basis (CB), is the set of
eigenfunctions of the enumerating operator $\hat{Z}:$

\begin{equation}\label{z}
\hat{Z}|n\rangle=\omega^n|n\rangle.
\end{equation}
 We shall designate this basis with b=$\ddot{0}$; i.e.
$|n\rangle=|n;\ddot{0}\rangle:$
Thus, the d+1 bases are  $b=\ddot{0},0,1,...d-1$.\\
We adopt the following abbreviation,
\cite{r2},$|\ddot{m}>=|\ddot{m};b=\ddot{0}>$ and $|m_0>=|m_0;b=0>$. We note,
for future reference, that the basis b=0 is made of the eigenfunctions of the
shift operator $\hat{X}$,
\begin{equation}\label{x}
\hat{X}|n>=|n+1>;\;|n+d>=|n>;\;\;\hat{X}|m;0>=\omega^m|m;0>.
\end{equation}
i.e. it is the Fourier transform of the CB. Note that the exponents are modular and may be viewed as
members of an algebraic field \cite{w1,w2,w3,r3}.\\
The URM we consider involves measuring an operator $\hat{K}$ of the general
form
\begin{equation}\label{K}
\hat{K}_b=\sum_m |m;b>\omega^m <b;m|;\;\;b=\ddot{0},0,1,...d-1,
\end{equation}
for some selected alignment (=basis), b. The URM considered is a measurement
of $\hat{K}_b$, Eq.(\ref{K}), of  particle 1 with an outcome m in a basis b
 with the values of m and b  unavailable, lost. Both the initially prepared
state and the control measurement basis
involve entangled systems: particle 1, the system subjected to the URM, and the ancilla, particle 2.\\
We shall show below that there are two "natural" control measurements that
provide distinct pieces of information. These are measurements of MUB of
maximally entangled states (MES) bases. (The MES considered here are pure two
particle states such that partial tracing over either one leaves as unity the
density matrix of the other.)  The presentation of these MES is simplest with
the use of collective coordinates which are now introduced schematically \cite{r1,r2,r3}.\\
The  Hilbert space of two d-dimensional particles, 1 and 2, is spanned by
$|n_1>|n_2>,\;n_i=0,1,...d-1,\;i=1,2$ where $|n_i>$ is the eigenfunction of
$\hat{Z}_i$, (i=1,2) corresponding to Eq.(\ref{z}), with similar relation for
the shifting operators, $\hat{X}_i$, viz $\hat{X}_i|n_i>=|n_i+1>$,
Eq.(\ref{x}). The space may, alternatively, be accounted for with collective
coordinate $|n_c>|n_r>$, $n_j=0,1,...d-1,\;j=c,r$. Here c stands for "center
of mass" and r for "relative" coordinates. These are defined for a
d-dimensional Hilbert space (d$\ne 2$) via the single particle dynamical
variables (recall, \cite{w3,r1}, that the exponents are modular variables,
e.g. $1/2= (d+1)/2 \;Mod[d]$), d=odd prime,
\begin{eqnarray}\label{coll}
\hat{Z}_c&=&\hat{Z}^{1/2}_1\hat{Z}^{1/2}_2;\;\;\hat{Z}_r=\hat{Z}^{1/2}_1\hat{Z}^{-1/2}_2;\;\;\hat{Z}_j|n_j>
=\omega^{n_j}|n_j>;\;n_j=0,1,...d-1.\;j=c,r\nonumber\\
\hat{X}_c&=&\hat{X}_1\hat{X}_2;\;\;\hat{X}_r=\hat{X}_1\hat{X}^{-1}_2;\;\;\hat{X}_j|n_i>=|(n+1)_j>;\;j=c,r; \nonumber \\
\hat{Z}_j\hat{X}_j&=&\omega
\hat{X}_j\hat{Z}_j;\;\;\hat{Z}_i\hat{X}_j=\hat{X}_j\hat{Z}_i;\;i\ne
j;\;i,j=c,r \;\;\hat{X}^d_j=\hat{Z}^d_j=I;\;j=c,r.
\end{eqnarray}
One may consider MUB for the c and r coordinates for the d+1 bases. We,
however, concern ourselves with the two collective coordinates MUB , the CB,
$b=\ddot{0}$, and its Fourier transform, b=0. We  label the CB bases of the
collective coordinates in close analogy to the particles ones. Thus the
"center of mass", c, CB basis, i.e. the eigenfunctions of $\hat{Z}_c$, are
$\{|n_c>\}\;=\;\{|n_c;\ddot{0}_c>\},\;n_c=0,1,...d-1,$. With similar
designation scheme for the eigenfunctions of $\hat{X}$, the b=0 case. (We
shall omit the basis subscript, e.g. $\ddot{0}_c\Rightarrow\ddot{0}$. Whenever
possible confusion may arise between values of b for the collective
coordinates with those of the
single particles it is removed via a detailed specification.)\\

Direct calculation proves \cite{r1,r4} that each $|n_1>|n_2>$ state
corresponds to a unique collective state $|n_c>|n_r>$:
\begin{equation}\label{colpar}
|n_1>|n_2>\Leftrightarrow|n_c>|n_r>;\;\text{with}\;
n_c=\frac{(n_1+n_2)}{2};\;\;n_r=\frac{(n_1-n_2)}{2}\Leftrightarrow
n_1=n_c+n_r;\;\;n_2=n_c-n_r.
\end{equation}
Adopting the following notational simplification for both c and r, \cite{r2},
viz.
$$|\ddot{n}>_i\equiv|\ddot{n};b=\ddot{0}>_i;\;\;|n_0>_i\equiv|n_0;b=0>_i;\;\;i=c,r \;,$$
we now prove that the product collective state $|\ddot{m}>_c|2m_0>_r$ is a
maximally entangled state (MES). Indeed
\begin{eqnarray}\label{cr}
|\ddot{m}>_c|2m_0>_r&=&|\ddot{m}>_c\big[\frac{1}{\sqrt
d}\sum_{n=0}^{d-1}|n>_r\omega^{-2m_0 n}\big]\nonumber \\
&=&\frac{1}{\sqrt
d}\sum_{n=0}^{d-1}|\ddot{m}+n>_1|\ddot{m}-n>_2\omega^{-2m_0n},
\end{eqnarray}
where we used Eq.(\ref{colpar}). The last expression is obviously a MES,
QED.\\
Now the $d^2$ MES: $|\ddot{m}>_c|2m_0>_r,\;\ddot{m},m_0=0,1,...d-1,$ are
orthonormal and span the two d-dimensional particles Hilbert space and thus
form a (MES) basis for it. This MES basis defines a conjugate basis made of
the $d^2$ MES: $|\ddot{m}>_r|2m_0>_c,\;\ddot{m},m_0=0,1,...d-1,$ (r and c are
interchanged). The two bases are MUB:
\begin{equation}
|<\ddot{m}{'}|_c<2m{'}_0|_r|\ddot{m}>_r|2m_0>_c|=\frac{1}{d};\;\text{independent
of}\;\ddot{m},\ddot{m}{'}, m_0,m{'}_0.
\end{equation}

Either basis may be used as a retrieving control measurement. Thus one control
measurement involves measuring the operator

\begin{equation}\label{gamma1}
\hat{\Gamma}^a=\sum_{\ddot{m},m_0}|\ddot{m}\rangle_c|2m_0;0\rangle_r\Gamma_{\ddot{m},m_0}^a
\langle \ddot{m}|_c\langle 0;2m_0|_r,
\end{equation}
This operator involves the double (commuting) collective operators,
Eq.(\ref{coll}). (This control measurement  relates the NSM issue to the so
called Mean King Problem \cite{v,bg2,r3}.) The other, conjugate control
measurement involves measuring
\begin{equation}\label{gamma2}
\hat{\Gamma}^b=\sum_{\ddot{m},m_0}|\ddot{m}\rangle_r|2m_0;0\rangle_c
\tilde{\Gamma}_{\ddot{m},m_0}^b \langle \ddot{m}|_r\langle 0;2m_0|_c,
\end{equation}
$\hat{\Gamma}^b$ involves the double (commuting) collective operators
conjugate to those of Eq.(\ref{gamma1}) it relates to the so called Tracking
the King problem
\cite{r3}.\\

\section{Measuring Unrecorded Measurement}

We now outline two  protocols wherein measuring a URM  allows  retrieval of
some of  unavailable data of a URM of a d-dimensional particle 1: Let Alice
prepare the MES $|\ddot{m}>_c|2m_0>_r$, \cite{r4,r5},wherein particle  1 and
an ancilla,
particle 2, are (maximally) entangled.\\
Now let Bob measure $\hat{K}_b$, Eq.(\ref{K}), with an outcome m. (The basis
(b) of the measurement and the outcome (m) are not available to Alice.) The
state of the two particle system is now (unnormalized)\cite{barnett}, using
 Eqs.(\ref{mub}),(\ref{colpar}),(\ref{cr}),
\begin{eqnarray}
|m;b>_1<b;m|\ddot{m}>_c|2m_0>_r&=&\frac{1}{\sqrt d}|m;b>_1|-m";-b>_2\omega^{\phi}, \nonumber \\
&&m"=2(m_0+b\ddot{m}-b/2)-m;\;\phi=
-\frac{b}{2}(2\ddot{m})(2\ddot{m}-1)+2\ddot{m}m.
\end{eqnarray}

Let Alice select, as her control measurement, to measure $\hat{\Gamma}^a$,
Eq.(\ref{gamma1}), having as her outcome, say,
$\Gamma_{\ddot{m}{'},m{'}_0}^a$. Thus she is assured that the following matrix
element is non-vanishing ($b\ne \ddot{0})$:

\begin{equation}
0\ne <\ddot{m}{'}|_c<2m{'}_0|_r|m;b>_1|-m';-b>_2\;\Rightarrow\; m'_0+b\ddot{m}{'}=m_0+b\ddot{m}.
\end{equation}
This implies (note: the equations are modular)
\begin{eqnarray}
b&=&\frac{(m_0-m{'}_0)}{(\ddot{m}{'}-\ddot{m})}\;\;\ddot{m}\ne \ddot{m}{'}. \nonumber \\
 &=&\ddot{0},\;\;\;m_0\ne m{'}_0,\;\;\;\ddot{m} = \ddot{m}{'}.\nonumber \\
 &=& \text{undetermined for}\;\;m_0= m{'}_0,\;\;and \;\;\ddot{m} = \ddot{m}{'}.
\end{eqnarray}
The last two results are obtained upon evaluating the matrix elements that
include $b=\ddot{0}$, i.e. the possibility of Bob  measuring
$\hat{K}_{(b=\ddot{0})}$.\\
The protocol above reveals the basis used in the URM except for the case
wherein the outcome of the control measurement recovers the initially prepared
state which would also be the case wherein no measurement, either the usual or
URM, were performed.
In such a case the outcome of the control measurement does not reveal any of the sought after data.\\
Equivalent results are gotten when Alice prepared state is the conjugate state: $|\ddot{m}>_r|2m_0>_c$  and
correspondingly, the control measurement is  $\hat{\Gamma}^b$, Eq.(\ref{gamma2}).\\
An alternative protocol for the unveiling of (some) of the data of a URM is
one wherein we proceed as above but replace the control measurement
$\hat{\Gamma}^a$, Eq.(\ref{gamma1}), with  $\hat{\Gamma}^b$,
Eq.(\ref{gamma2}). In this case an outcome of the control measurement,
$\Gamma_{\ddot{m}{'},m{'}_0}^b$, assures the non vanishing of

\begin{equation}
0\ne <\ddot{m}{'}|_r<2m{'}_0|_c|m;b>_1|-m';-b>_2\;\Rightarrow\; m-2(m{'}_0+b\ddot{m}{'})=2(m_0+b\ddot{m}-b/2)-m.
\end{equation}

This implies,
\begin{equation}
m=(m_0+m{'}_0)+b(\ddot{m}+\ddot{m}{'})-b/2.
\end{equation}

i.e. this protocol unveils a relation between the basis, b, and the outcome,
m, of the URM. (Note: this case is closely related to the so called Mean King
Problem \cite{v,bg1,bg2,r2}: it allows the deduction of the
outcome, m, given the basis used, b.\\

\section{Conclusions and Remarks}

An Unrecorded Measurement (URM) is a projective (Von Neumann) measurement of
an operator, selected from a prescribed complete set of operators, with the
measurement outcome and the operator selected unavailable, "lost". Such a
measurement is completely inconsequential within classical physics (CP) since
within CP a measurement leaves the measured system essentially unperturbed.
Within quantum mechanics (QM) the measurement marks the measured system. Two
distinct protocols were considered in the present work, each involving a
distinct control measurement
whose outcome reveals some of the "lost" data of a URM.\\
The protocol prescribes a state preparation wherein the particle to be
measured without record is entangled with an ancilla. The correlations
involved in the entanglement reveal, via the outcome of the respective
control measurement, the sought after data.\\
The present study is confined to d-dimensional Hilbert space particles with d
an odd prime as for these dimensionalities complete sets of mutual unbiased
bases (MUB) are known. The extension of the theory to d being a power of prime
is possible but is judged to require complicated mathematics without adding
physical insight. The case of d=2 requires a special treatment which is left
for future work.\\

  The essential role played by entanglement in the unveiling change of the
quantal state due to (projective) measurement relates to the intimate relation
among entanglement, measurement theory and thence to the uncertainty
principle.

\end{document}